# OVERLOOKED ASTROPHYSICAL SIGNATURES OF AXION(-LIKE) PARTICLES

Konstantin ZIOUTAS [1,2], Mary TSAGRI [1], Yannis SEMERTZIDIS [1,3],
Thomas PAPAEVANGELOU [4], Annika NORDT [5], Vassilis ANASTASSOPOULOS [1]

1) University of Patras, Patras, Greece
2) CERN, Geneva, Switzerland
3) Brookhaven National Laboratory, USA
4) DAPNIA, CEA-SACLAY, Gif-sur-Yvette, France
5) TU-Darmstadt & MPE-Garching, Germany

Emails: zioutas@physics.upatras.gr ; mary.tsagri@cern.ch ; yannis@bnl.gov ;
thomas.papaevangelou@cern.ch ; anordt@cern.ch ; vassilis@upatras.gr

**ABSTRACT**

We discuss solar signatures suggesting axion(-like) particles. The working principle of axion helioscopes can be behind unexpected solar X-ray emission, even above 3.5 keV from non-flaring active regions. Because this is associated with solar magnetic fields ($\sim B^2$), which become in this framework the catalyst and not the otherwise suspected / unspecified energy source of solar X-rays. In addition, the built–in fine tuning we may (not) be able to fully reconstruct, and, we may (not?) be able to copy. Solar axion signals are transient X-ray brightenings, or, continuous radiation from the corona violating the second law of thermodynamics and Planck's law of black body radiation. To understand the corona problem and other mysteries like flares, sunspots, etc., we arrive at two exotica: a) trapped, radiatively decaying, massive axions allow a continuous self-irradiation of the Sun, explaining the sudden temperature inversion ~2000 km above the surface and b) outstreaming light axions interact with local fields ($\sim B^2$), depending crucially on the plasma frequency which must match the axion rest mass, explaining the otherwise unpredictable transient, but also continuous, solar phenomena. Then, the photon energy of a related phenomenon might point at the birth place of involved axions. For example, this suggests that the ~2 MK solar corona has its axion roots at the top of the radiative zone. The predicted B ≈ 10–50 T make this place a coherent axion source, while the multiple photon scattering enhances the photon-to-axion conversion unilaterally, since axions escape. We conclude that the energy range below some 100 eV is a window of opportunity for axion searches, and that it coincides with a) the derived photon energies for an external self-irradiation of the Sun, which has to penetrate until the transition region, and b) with the bulk of the soft solar X-ray luminosity of unknown origin. Thus, (in)direct signatures support axions or the like as an explanation of enigmatic behavior in the Sun and beyond; e.g., the otherwise unexplained "solar oxygen crisis" taking into account related observations ($\sim B^2$) in pores, which is associated with X-ray emission. Axion antennas could take advantage of such a feed back. Finally, the observed soft X-ray emission from the quiet Sun at highest latitudes as well as the extended activity associated with magnetic structures crossing the solar disk centre suggest that a multi-component axion(-like) scenario is at work.

## 1. Introduction

With the ongoing direct search for (solar) axions [1] one can ask whether the same detection principle is at work behind certain mysterious solar or other astrophysical observations. Indirect signatures for dark matter particle candidates might be strong, if, for example, the same process(es) can explain consistently more celestial phenomena. Here we mainly discuss striking solar observations of unknown origin which seem to require the involvement of axions or other exotica with similar properties. Previous work [2–4] addressed a steady X-ray emission due to very few gravitationally trapped massive axion-like particles (about 1 in $10^7$, or only ~100 *kg*/s since 4.6 Gyears), giving rise to a self-irradiation of the whole Sun. In this work, we mainly argue in favour of a second solar X-ray component expected to originate from the widely accepted interaction of light axions or the like with the inner/outer solar magnetic fields, as they stream out of the Sun. In this work we emphasize the ubiquitous solar magnetic field, which is the dominating "element" in all axion experiments.

## 2. Astrophysical signatures

***The solar corona problem:*** Stellar observations and theory of stellar evolution cannot be reconciled with stars having atmospheres that emit X-rays [5], suggesting the question: Where do these X-rays come from? The mechanism that heats the solar corona to some MK has remained elusive since its discovery in 1939. The solar corona problem is one of the most important and challenging problems in astrophysics [6], since it violates, at first sight, the second law of thermodynamics, which is actually improbable. The radiative decay of gravitationally trapped massive particles like the generic Kaluza–Klein axions, being created by the Sun itself, could provide the invisible source that sustains the self-heating of the solar atmosphere [2], thus reconciling observation and thermodynamics. In fact, the expected decay hard X-rays from massive axions, accumulated around the Sun since cosmic times, fit only the energetic part (above ~2 keV) of the reconstructed analog photon spectrum from the quiet Sun during solar minimum (see Figure 8 and 9 in ref. [2]). However, this is only one X-ray component within the axion scenario, and probably the weakest one.

Interestingly, it was shown in ref. [3] that the radial distribution of the two off-pointing observations with the YOHKOH X-ray satellite fit the massive axion scenario also in the low–energy range (~0.5–4 keV), thus suggesting more measurements of this kind. In fact, the RHESSI mission has recently repeated several off-pointing observations above the solar limb during non-flaring, spotless and active-region-free Sun, i.e., during quiet Sun conditions [7], arriving apparently at a hard X-ray spectrum, which is rather similar to the reconstructed one from YOHKOH measurements in the previous solar cycle minimum (see Figure 9 in ref. [2]). The values obtained with RHESSI in the ~3–12 keV range, for such extremely quiet periods at the present solar minimum phase, correlate with those from another orbiting X-ray solar telescope (GOES), implying a real signal in this actually not so low energy band [7,8] for a cool Star like our Sun. This we consider as additional supporting evidence for the (massive) solar axion scenario [2]. Because, a conventional explanation with electrons raises a new problem: How is such a population of energetic electrons created in the quiet corona?

***Magnetic field related observations:*** The long lived massive axion(-like) particles can not explain:
- a) local and transient solar phenomena, and
- b) the prevailing soft X-ray emission, which makes the bulk of the solar X-ray luminosity during periods of quiet (see Figure 9 in ref. [2]) as well as active [i.e. flaring] Sun.

Therefore, an as yet unnoticed additional X-ray source must be at work. The Primakoff–effect with "conventional" axions, e.g. like the QCD-motivated ones, can interact inside the solar magnetic fields, which is after all the most natural process to expect, as it resembles the working principle of an axion helioscope like CAST. Thus, near the solar surface X-rays should also be emitted due to magnetic-field-induced radiative "decay" of outstreaming axions (see below). Depending on the relative intensity of converted light axions, the resulting radial distribution of X-rays, coming from both components near the surface of the Sun, can be different from that expected for massive ones only [2,3]. Interestingly, the available two off-pointing observations with YOHKOH [3] show a radial distribution that agrees within ~30% with the massive axion simulation. Therefore, in our heuristic approach explaining the otherwise unexpected solar X-rays with axions or the like, this second magnetic–field–related component offers a possible explanation also for this rather small but significant discrepancy.

There is strong observational evidence that (transient) solar X-ray emission correlates with the local magnetic field strength squared (~$B^2$) [9], which is characteristic for light axion involvement, and, it determines the performance of an axion helioscope [1] à la Pierre Sikivie, and, à la Karl van Bibber *et al.*. Actually in all axion experiments, the $B^2$ - dependence of a potential signal is generally accepted as the ultimate method for axion identification. Large–scale magnetic fields of several kGauss exist in the enigmatic sunspots, which show enhanced and unpredictable activity. Interestingly, it is known that the magnetic field somehow heats the quiet solar corona, with the exact energy release mechanism still enigmatic in solar physics; within the axion picture, this is naturally expected. A few magnetic–field–related cases in favour of the axion(-like) scenario have already been discussed in ref. [9]. For completeness, we repeat some of them here briefly in an updated form (see also ref. [9]):

a) During the last solar cycle minimum in 1996, the X-ray emission from an isolated solar active region, measured with YOHKOH's X-ray telescope, during non-flaring periods, shows a striking $B^{1.94\pm0.12}$ dependence in the range ~0.5–4 keV [41]; within the axion scenario, this points at a low energy solar axion spectrum (below ~ 1 keV), in contrast with the widely used solar axion spectrum that peaks at ~4.2 keV. Such a new component is of utmost interest for the design of (earth bound) direct solar axion searches (see also additional observations given below in section 4). Similarly, a plethora of solar soft X-ray flux measurements correlates with the local magnetic field (~$B^n$), with the exponent n≈2 varying smoothly by ~20% during the solar cycle [10], pointing possibly to a deeper axion implication in the dynamic Sun. Furthermore, Wolfson, Roald, Sturrock & Weber [39] investigated a data set of 521 days beginning 1996 July 25, and, they find strong spatiotemporal correlations between soft X-ray emission and the underlying magnetic field at all but extreme latitudes. The derived relation between photospheric magnetic field and X-ray flux with a power-law index of ~1.87 is close to the $B^2$-dependence. Remarkably, no correlation was found with the line-of-sight magnetic field component (as it is expected for an axion-photon conversion). Moreover, this correlation becomes near negligible at the highest latitudes considered in [39]. Interestingly, the authors of ref. [39] conclude: "these correlations hint at a relation between the strength of the photospheric field and the heating processes that ultimately result in coronal X-ray emission although they provide no insight into the nature of such a relation".

b) Remarkably, above sunspots the corona is hotter (e.g. ~2.1 MK instead of 1.3 MK) and the photosphere just underneath is cooler (e.g. ~4000 K instead of 5770 K) than near quiet Sun regions.
Each of these observations is consistent with a Primakoff–effect taking place inside the extended strong surface magnetic fields with the plasma frequency matching the axion rest mass. The emerging picture is: 1) energetic axions streaming out of the hot inner Sun can be converted back to X-rays, further heating–up the preexisting ~1.3 MK quiet Sun corona, and 2) near the photosphere or even deeper into the Sun, photons with energy below ~10–100 eV can undergo the reverse Primakoff–process escaping into axions or the like, thus making the photosphere cooler.

c) Generally, there is strong evidence that magnetic elements in the Sun with higher magnetic flux are less bright [11]. Low–energy solar axion production due to surface magnetic fields fits the observed ~$B^2$-dependent dimming effects of the sunspot brightness in the visible. Also recently [12], some 900 sunspots have shown an increase in brightness while the magnetic field strength was decreasing. Then, the dark sunspots can be low–energy axion sources, whose strength should correlate with their level of darkness, thus providing an axion trigger [13]. To be more quantitative, we work out here a ***numerical example***:   We use the PVLAS derived parameters, i.e. $g_{a\gamma\gamma} \approx 2.5\ 10^{-6}\ \text{GeV}^{-1}$ and $m \approx 1$ meV. We take as a mean free path length of visible photons in the photosphere equal to ~100 km, which is equal to the coherence length of the photon-to-axion conversion; the condition $\hbar\omega_{plasma} \approx mc^2 \approx 1$ meV seems reasonable. For a sunspot magnetic field $B \approx 3$ kGauss (see ref. [21]), and, provided the PVLAS finding is correct, we estimate a 'photon-to-axion' conversion efficiency $P_{\gamma a} \approx 0.001$. Furthermore, assuming a surface of a single sunspot of a few ‰ of the solar surface, the expected solar axion-like luminosity is $L_{axion} \approx 10^{28}$ erg/s. The derived signal rate in the visible, applying CAST performance [1], is $S \approx 300$ Hz. Such a high signal rate along with the rather conservative values we used, allow to be sensitive to a much smaller coupling constant and check the PVLAS result.

In summary, axion–photon oscillations depend a) on the squared transverse magnetic field component along the axion/photon propagation, b) on the local plasma density, since at resonance $\hbar\omega_{plasma} \approx m_{axion} c^2$ the coherence length is limited by the photon absorption, and c) on the magnetic field configuration. Note that each of these parameters changes permanently near the solar surface and also deeper inside the turbulent (i.e. boiling) convective zone. For a quantitative calculation, since it implies restless density fluctuations for any place above the radiative zone ($R > 0.7 \cdot R_{solar}$), it is crucial to know this dynamical behaviour; resonance can be restored temporally repeatedly covering a large bandwidth in axion rest mass, e.g., ~0-10 eV/$c^2$, if we take as maximum density the one near the bottom of the convective zone. Then, axion (dis)appearance is not at all a static process; it can be the cause of the unpredictable dynamical behaviour of the Sun from the visible to X-rays. In fact, such a rapid change of the axion-relevant parameters defines instantly certain volume elements, where resonance-coherence effects can result in an enhanced axion production as well as in axion-related brightenings or light dimming. Note that a fine tuning is also exercised in CAST Phase II, by changing the density of the refractive gas in the magnetic volume [1].

Flares: What produces solar flares and Coronal Mass Ejections is one of the great solar mysteries. However, the energy that powers them is generally believed to be (connected to) the magnetic field [14]. Following recent work [15], these strong X-ray emitting events correlate (within 1.8$\sigma$) with the solar surface magnetic field-squared (see Figure 1). In fact, our fit to the X-ray data gives a ~$B^{1.6 \pm 0.22}$ dependence, while the authors in ref. [15] refer even to a $B^2$ correlation, fitting the assumed axion scenario from ref. [15]. The blue dashed line gives the derived fit to the data points (~$B^{1.6 \pm 0.22}$). Note that the conventional solution for this radiation emission is that the hard X-rays are coming from electrons. However, this leads to the electron "problem" [16], since the energetic electron flux must be ~$10^5$ times higher than the X-rays. We stress here that within the axion scenario the local surface magnetic field is

"only" the required catalyst for the axion–photon reactions to take place, and not the otherwise suspected / unspecified energy source of solar X-rays. In this framework, the inner Sun is the actual energy source that creates the outstreaming axions.

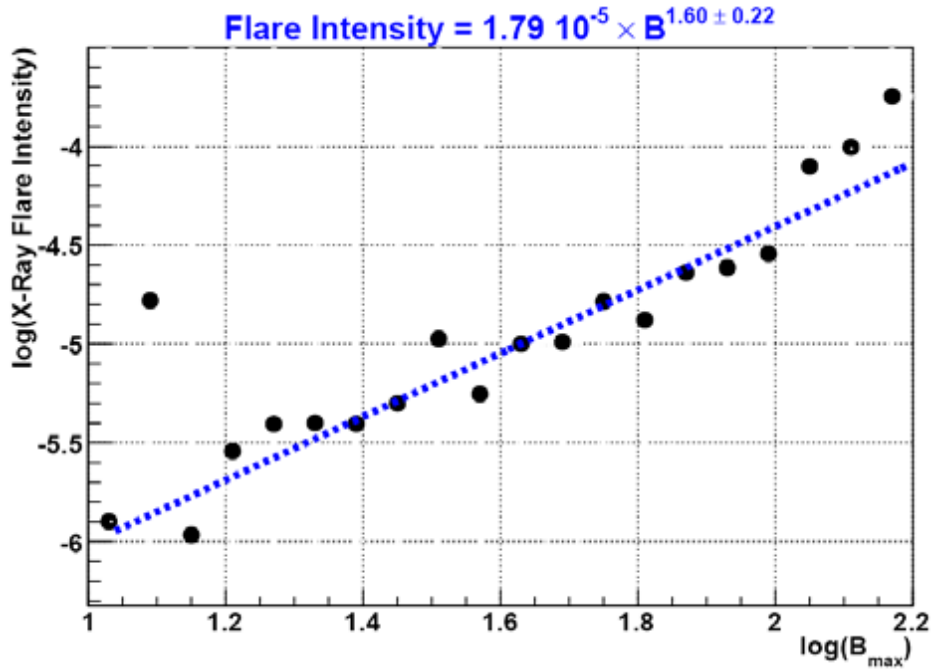

**Figure 1** Rebinned peak flare X-ray intensity vs. the maximum magnetic field ($B_{max}$), reconstructed

## 3. Solar magnetic fields

We recall that it is the macroscopic coherence effects in magnetic axion helioscopes [1] which result in an enhanced axion-to-photon conversion than in detectors, where the axions interact incoherently with the detector atoms. Therefore, helioscopes are more sensitive to axions. Surprisingly, solar magnetic fields [17] have not been included in the model calculations of the axion production rate. However, inside the magnetic Sun, when the local plasma frequency fits the axion rest mass, photon–axion conversion might even be the dominant axion production channel, since the coherence length can be maximum, i.e., equal to the photon mean free path. This is ~1 mm in the core and ~100 km (!) at the photosphere. Such an additional solar axion production channel can modify the previously expected solar axion spectrum (its intensity and its shape) without taken into consideration magnetic fields [18]. To know this is of utmost importance in interpreting solar observations as well as in adopting the appropriate parameter values for optimum performance of an axion antenna.

Interestingly, strong magnetic fields outside the hot core, like the simulated ones in Figure 2, can selectively enhance the production of low(est) energy axions. More precisely, in this scenario, the soft quiet Sun X-ray luminosity (from the celebrated ~2 MK corona) could originate from converted massive or light axions due to their spontaneous or "induced" decays, respectively, both kinds of axions stemming from a shell between the radiative and the convective zone (R ~ 0.7 $R_{solar}$); since it has the required temperature of ~2 MK [19] and is permeated with a predicted ~ 30-50 Tesla magnetic field (see Figure 2), this is a potential source of low–energy axions. The same reasoning applies to other potential candidates, e.g. microflares with a mean temperature of 12.6 MK [20], including large flares whose temperature distribution is, remarkably, around ~15-20 MK. It is also interesting to note that the power of flares, extrapolated on the whole solar surface, does not actually exceed the total solar luminosity [20]. This is not a signal, but it fits with the heuristic picture we follow with flares too, while in the opposite case, the reasoning would have been negative.

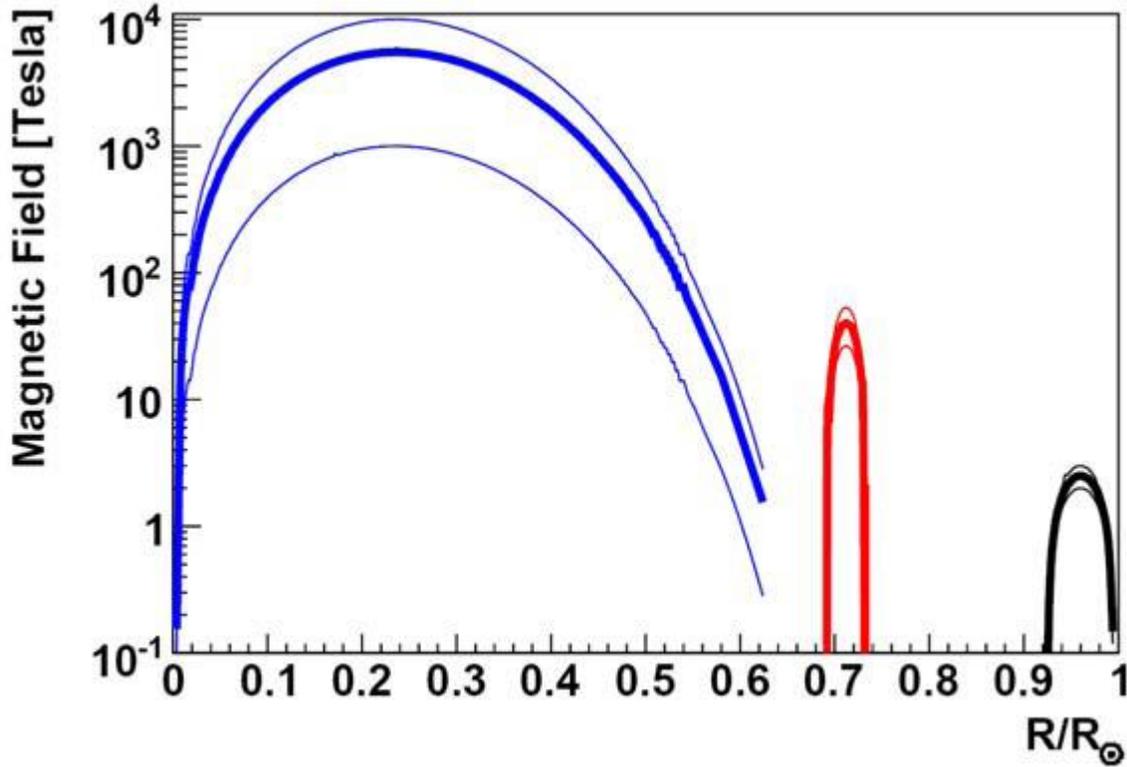

**Figure 2**  The reconstructed solar magnetic field simulation from ref. [17]: $10^3$–$10^4$ Tesla (left), 30–50 Tesla (middle) and 2–3 Tesla (right), with a temperature of ~9 MK, ~2 MK and ~200 kK, respectively. The thin lines show the estimated range of values for each magnetic field component. Internal rotation was not included in the calculation. An additional axion production at those places can modify both intensity and shape of the solar axion spectrum (Courtesy Sylvaine Turck-Chièze).

We stress that a few 100 eV massive axions of the Kaluza–Klein type might also be created coherently, since the many axion mass states allow a quasi continuous resonance crossing throughout the magnetic Sun, which is a unique fine tuning, only if Kaluza–Klein-like axions exist. If these less massive axions (see ref. [2]) build up a large trapped component around the Sun, their spontaneous decay could explain the low–energy X-rays from the quiet Sun. The induced "decay" in magnetic fields of light axions at the same lower energy range, created probably in the same place, comes in addition. In favour of this conclusion we recall the following findings:

 a) the quiet Sun soft X-ray intensity is increasing at lower energies and its origin
    is unknown,
 b) the measured radial distribution of the two off-pointing observations with
    YOHKOH (~0.5–4 keV) actually supports such a less massive axion scenario,
    which was not covered in ref. [2],
 c) in order to reach the depth of the solar transition region (at ~2000 km above

the photosphere) of the Sun being externally self-irradiated from decaying axions orbiting around the Sun, the estimated photon energies were ~50–350 eV [2], which coincides with the energy range of the ~2 MK solar corona radiation. So far, the estimated X-ray luminosity due to the generic massive solar axions was even strongly suppressed below ~1 keV [2], since their assumed birth place was the ~16 MK hot solar core. Therefore, it seems that magnetic fields might be one of the key parameters for the solar axion scenario that can accommodate light as well as massive axions or the like.

Furthermore, local magnetic fields up to ~0.6 Tesla have been measured at the surface of sunspots [21], where axion-to-photon oscillations and vice versa can take place and cause otherwise unexplained phenomena. Remarkably, already in 1983 Schadee-deJager-Svestka [35] observed "enhanced hard X-ray emission (above 3.5 keV) in non-flaring active regions … requiring coronal temperatures in excess of 10 MK". Then, the whole Sun is a multi-component/faceted axion source.

## 4. Soft X-ray emission from quiet Sun and active regions crossing solar disk.

In this section we present solar observations made with the Yohkoh mission ~10±1 years ago. They resemble features of actually direct signatures of solar axions or other particles with similar properties, according to our present interpretation. Because, some solar X-ray emission correlates with magnetic regions (an actually well known fact) showing also the expected spatiotemporal behaviour like the characteristic $B^2$-dependence of axion-photon conversion (Primakoff effect). We conclude here about the properties the actual solar axion source might have. In fact, it was proposed in ref. [36] that outstreaming solar axions could be converted to radially emitted X-rays in the magnetic field of a sunspot. The only place to directly observe, in this way, solar axions (with a broad energy distribution peaking at 4.2 keV) was expected to coincide with a sunspot crossing (~2 days) of the solar disk centre on the Earth-side, i.e., the detection principle of an axion helioscope (like CAST phase I and/or phase II [1]) could be at work there, and this almost cost free. However, following our evaluation of archived Yohkoh data as well as the re-evaluation of published Yohkoh results, in order to reconcile prediction with observation, we are led to assume instead:
  a) soft axions or (most probably) axion-like particles in the ~sub-keV range (see also **c)**),
  b) their place of birth extends eventually out to the tachocline at ~0.7$R_{solar}$
      (and probably beyond).
  c) the so far ignored inner magnetic fields of the Sun giving eventually rise to the possible involvement also of other exotica like millicharged particles, etc., which are expected to interact very feebly with ordinary matter.
      Their coupling to the magnetic field can result to an accumulation over cosmic times, without the need to invoke the usually inefficient gravitational (self)trapping [2].

In the following, we give a few solar observations, which we have reproduced from published work or derived from public data of the Yohkoh mission. It seems that an axion(-like) involvement in the Sun has a multifaceted appearance making their identification difficult, explaining why such signatures have been overlooked so far. Thus:

**A)** In May 1998, the Yohkoh XRTelescope has observed soft X-rays from an isolated small spotless active region outside flaring times, with a factor ~100 times less flux than classical active regions. It is correlated in time with the measured total magnetic flux [37]. Figure 3a shows this tiny sunspot in a 2D solar image in soft X-rays. In Figure 3b we also have reconstructed the time evolution of the smoothed X-ray emission (blue line) while crossing very close the solar disk centre, i.e., 1.8°N [38]. The duration of the X-ray emission (~2.5 days) and the crossing time fit the scenario of converted axions at the solar surface (near the disk centre) originating from the solar core. By contrast, the observed soft X-rays imply low energy axions or the like, which are actually suppressed in the conventional solar axion model. Otherwise, the spatiotemporal behaviour of this solar observation is significant and fits a quite revised axion scenario. Therefore, more similar solar bright points can exclude the random occurrence of this one (see Figure 5).

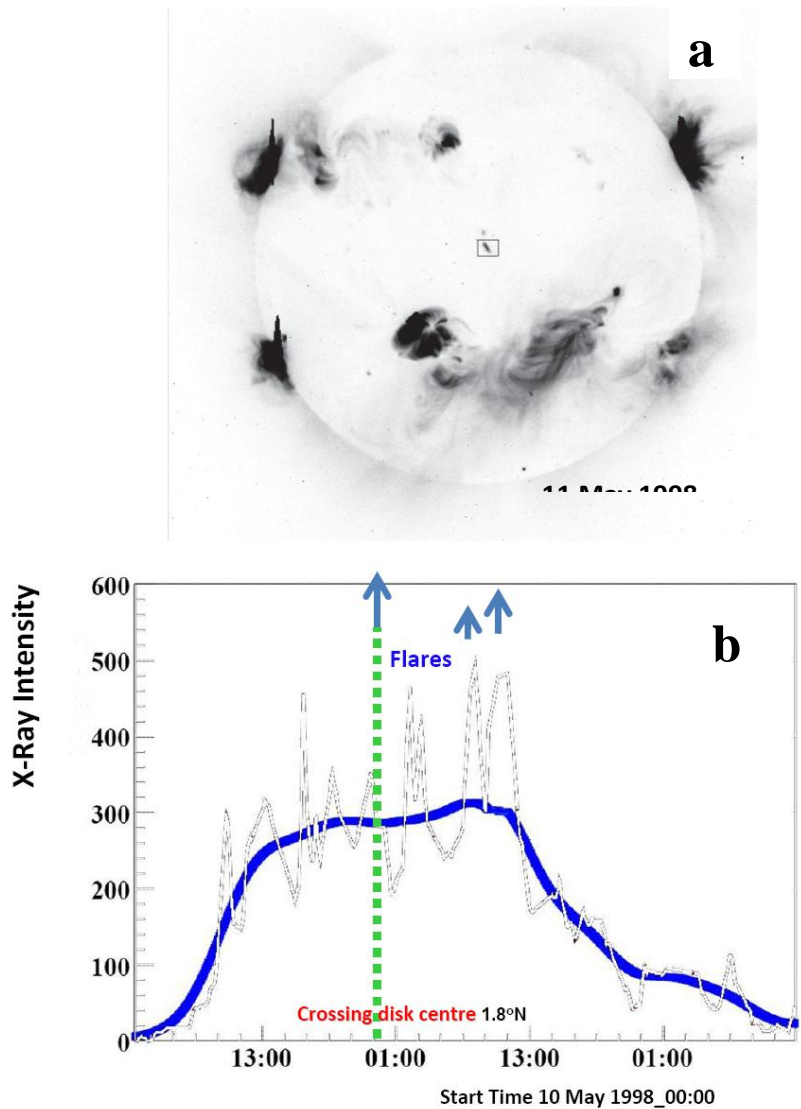

**Figure 3** (**a**) A tiny isolated sunspot in a 2D solar image in soft X-rays observed with the X-ray telescope of the Yohkoh mission in May 1998 [37]. It is well separated from other active regions by at least half solar radius. (**b**) Its full evolution (~2.5 days) from birth to disappearance occurred while crossing very close the solar disk centre at an latitude of 1.8°N [38]. The smooth thick blue line is derived from the measured points (dashed line), with the 3 flares not included in the smoothing process. The green dashed line shows the crossing time of the disk centre.

**B)** Pevtson & Acton [40] also analyzed Yohkoh XRTelescope data. The X-ray emission at the heliographic centre correlates with the magnetic flux in the period 1991 (active Sun) to 1996 (quiet Sun). For the analysis of X-ray irradiance of quiet Sun associated with the disk passage of an isolated active region (NOAA AR7981), they used three bins, each 4°x4° in size, centered at the solar disk centre, at 50°N, at 50°S and compute the temporal variation of the X-ray flux for each of these areas during three solar rotations (~2.5 months) starting July 23$^{rd}$ 1996, near solar minimum. Remarkably, a strong activity was measured each time this active region was crossing the central meridian. In Figure 4 (*top*) we have smoothed the time variation of the X-ray irradiance in these three quiet Sun areas during three consecutive solar rotations using Figure 7 of ref. [40]. Interestingly, the X-ray brightness at the high latitudes (some 100° apart) shows clearly remnants of low-latitude soft X-rays when the active region is near disk centre (i.e. in the central meridian). This requires even a more revised axion-like picture, i.e. an extended inner solar source of low-energy axion(-like) particles. To make the extended size of the solar X-ray emission more visible than it is actually seen in Figure 4 (*top*) every time the active region crosses the disk centre, we calculated the correlated spectrum of all three individual measurements. Figure 4 (*bottom*) is derived by multiplying all three simultaneous X-ray flux values taken at the equator, at 50°N, at 50°S, and smoothing the obtained spectrum. In this way we arrive at a curve (Figure 4 (*bottom*)), which shows more clearly the 3 main peaks plus a small one next to the peak at the middle. Then the X-ray activity associated with the crossing of the active region the central meridian is apparently coherent over such an extended region (~1.5 solar radii apart). For our axion-inspired interpretation this finding might point at a strong concentrated low energy axion source *towards* the solar core plus an additional extended one towards large distances from the solar core too. It is worth stressing here that taking into account the B$^2$-dependence of the X-ray emission from the same single active region present during solar minimum in 1996 for several months, along with the (repeated) brightenings maxima [37-40] across ~100° over the solar surface just during solar disk centre crossing, they can support only a revised solar axion(-like) scenario.

With luck, we may witness one or even more such events (like Figure 3) during the present solar cycle minimum too. For example, in Figure 5 we show the time evolution of the active region NOAA-AR10975, which crossed very near (=2°N) the solar disk centre. The arrows in each image point each time to the place of AR10975. During disk centre crossing an enhanced soft X-ray flux, as measured by the HINODE / XRTelescope, is apparent. Therefore, this seems at present to be of potential interest and it should be followed further.

**C)** Inspired actually from all these investigations, we also looked into the 2D distribution of solar X-rays, which was measured with the Yohkoh XRTelescope. Figure 6 compares 49 accumulated solar images during solar low and high activity. The soft X-ray emission of the active Sun is confined within ±45° in latitude as it is expected. The corresponding 2D plot for the quiet Sun at the solar minimum is actually the opposite as it results to a kind of clustering of events preferentially at high(est) latitudes; this we find unexpected, since magnetic activity, whatever its origin, is weaker towards the poles of the Sun, with ~95% of magnetic flux being within ±45° around the equator [42]. However, following the above concluded large extension of solar activity (along with the assumed generic inner solar axion(-like) source), this finding for the quiet Sun seems to be somehow reasonable, whatever the origin of these soft X-rays. Nevertheless, it might point at another origin/component than the enhanced X-ray observations given above from the disk centre during active region crossing.

Therefore, we are inclined to the conclusion that a novel solar axion(-like) scenario might be at work, which covers (almost) the whole Sun, being in addition free of all the "constraints" dictated by the conventional solar axion scenario. More theoretical and experimental work can shed more light on this kind of observations. It seems, however, that a generic axion-like scenario, with more than only one single component, might have to be invoked to explain these and other solar observations. This might be the reason why so far such otherwise potential strong signatures have been overlooked.

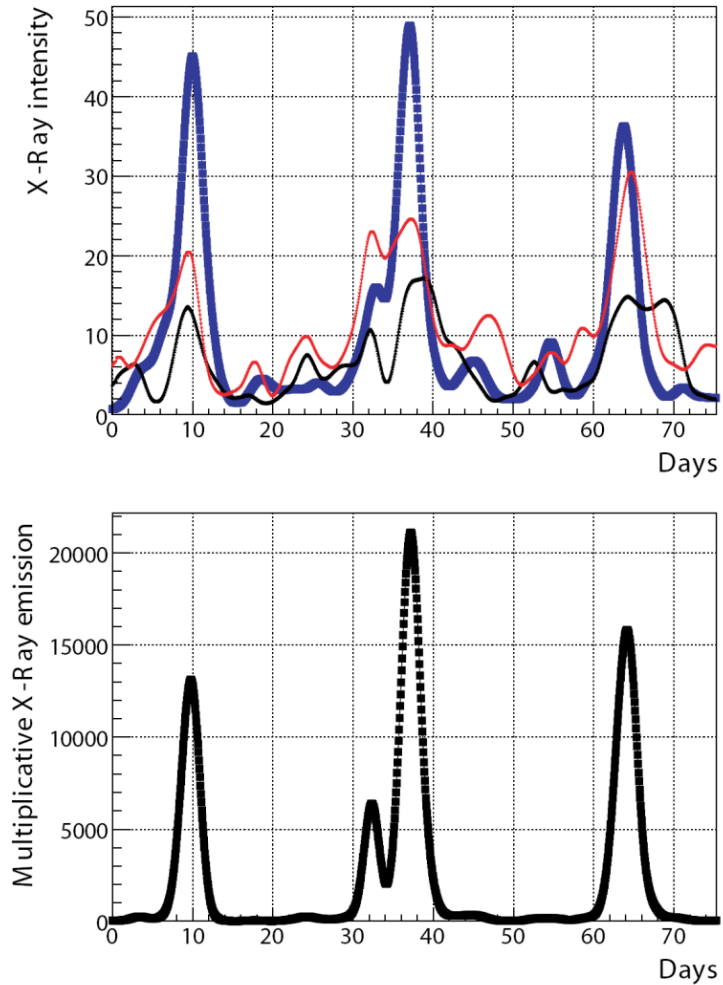

**Figure 4** (*top*) smoothed X-ray emission from quiet Sun areas located at the equator (*blue*), 50°N (*black*), 50°S (*red*) during 3 consecutive rotations associated with the solar disk passage of the single and isolated active region AR7981 (see Figure 7 in [40]). Start Time 23$^{rd}$ July 1996, 21:06:56. The temporally correlated X-ray emission from all three areas some 50° to 100° apart reduces the random events; the duration of each peak is a few days (*bottom*), pointing at extended and persistent process(es) at the Sun. The position in time of the main 3 maxima coincides strikingly with crossing the solar disk centre by the AR7981. A generic axion(-like) scenario implies an extended inner solar source of exotic particles with properties similar to the celebrated axions and a transverse magnetic field near the surface of the Sun, where the X-rays originate from the axion-to-photon conversion.

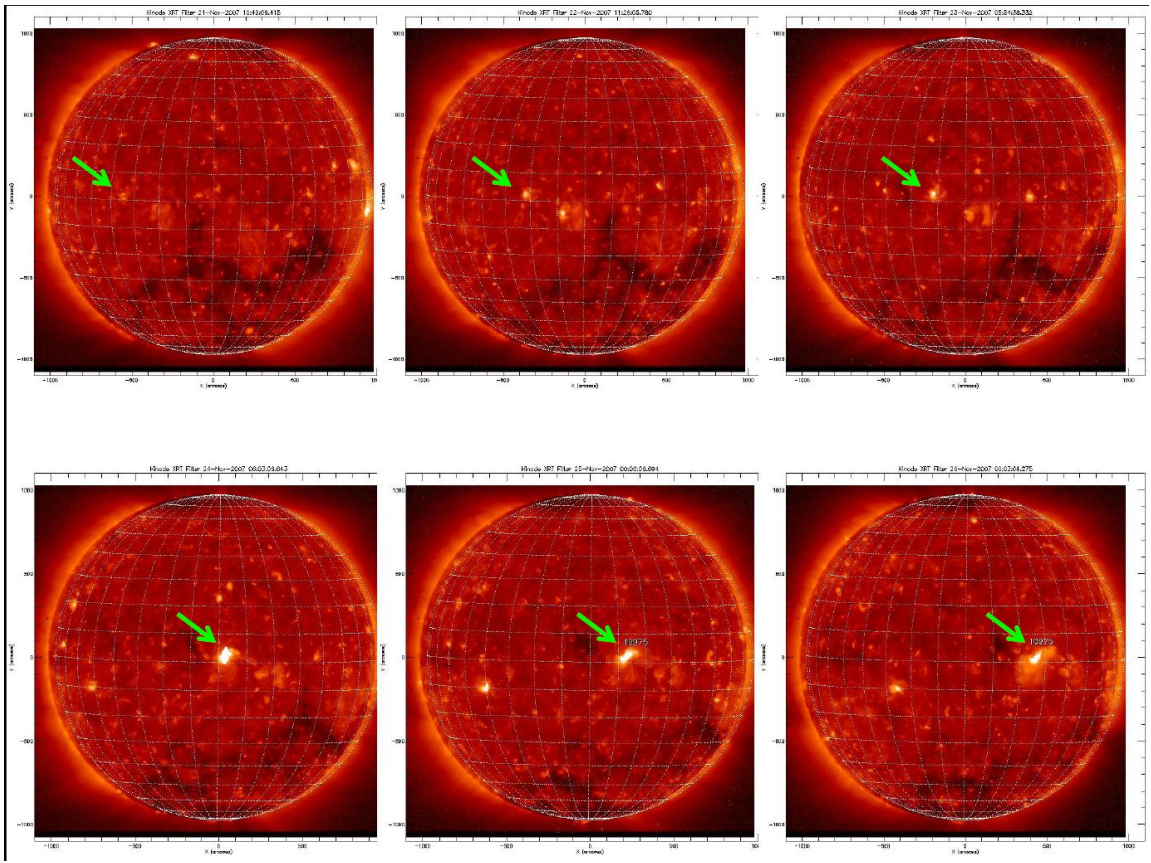

**Figure 5.** 2D full solar disk images in soft X-rays taken with the HINODE / XRTelescope in the period from 21$^{st}$ to 26$^{th}$ November 2007. The arrow points at the position of the AR10975. The left image on the bottom shows its passage near the disk centre (2$^o$N,2$^o$W)
(see also http://xrt.cfa.harvard.edu/data/latestimg.php ).

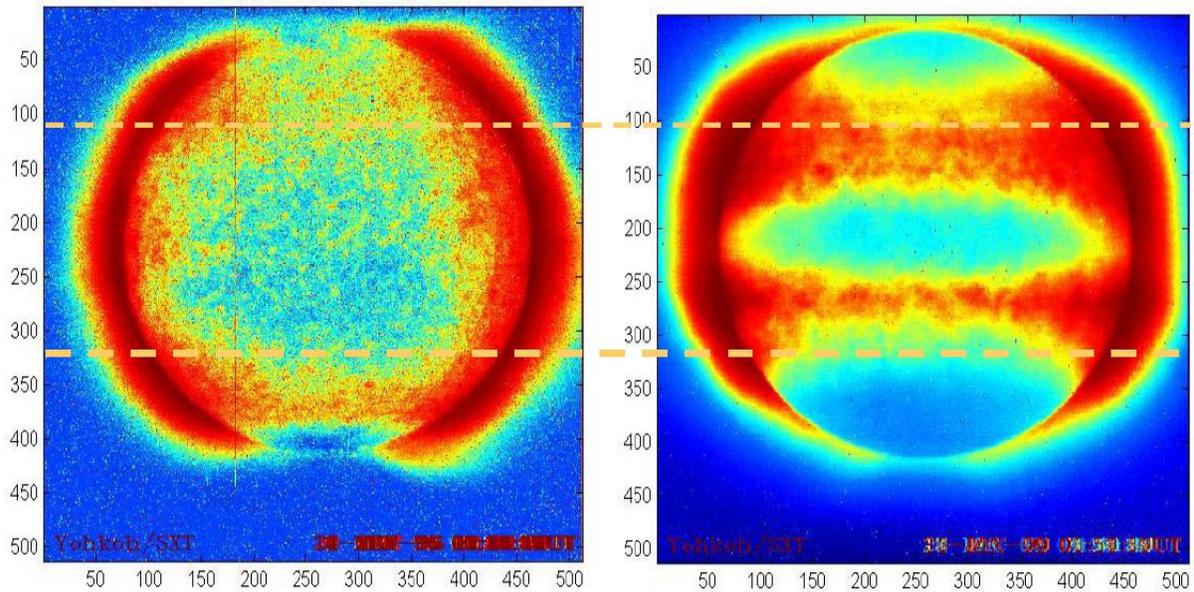

**Figure 6.** (*left*) Overlaid soft X-ray images from 49 quiet sun days taken with the Yohkoh XRTelescope in 1996 (solar minimum). The two horizontal dashed lines at $\pm 45^o$ show the region across the equator, where most of the solar activity occurs; this is demonstrated by the similar 2D plot (*right*) for the active Sun (solar maximum). Note the enhanced accumulation of events seen for the quiet Sun at high latitudes compared to the reduced one at low latitudes (*left*).

## 5. Discussion - Conclusion

Challenging questions like the origin of the soft and hard solar X-ray emission remain elusive within conventional astrophysics, with the solar corona problem being present ~70 years. Also, it has been known for many years that the magnetic field plays a crucial role in heating the solar corona, with the exact energy storage and release mechanism(s) being still a nagging unsolved problem for solar physics [22]. In this work, supporting evidence has been presented in favour of a second, magnetic–field–related solar X-ray component, which can originate from converted axions streaming out of the Sun, explaining thus also transient solar X-ray emission.

We also mention briefly other unpredictable X-ray observations, which seem to be of potential relevance for axions or the like: class 0 protostars [23]; the galactic centre, the Inter Cluster Medium, the ubiquitous X-ray background radiation [4]; the Fourier analyzed data of solar neutrinos and soft X-rays result in surprisingly identical frequencies [24].

The solar metallicity problem [25], i.e., the disagreement of the Ne/O abundance between solar model prediction and observation, could be due to an axion surface effect. Following Martin Asplund, a pioneer of this discovery, "*one possibility is rather the extra heating from the X-ray absorption of converted axions in the atmosphere which might increase the temperature in the spectral line formation region of the Sun*"[26].

Furthermore, the recently labeled "solar oxygen crisis" [27] along with the interesting observations performed by Socas-Navarro & Norton [27] seem to be of particular interest, as they fit qualitatively and quantitatively the advocated solar axion scenario. Our picture is this: if more X-rays are produced near the solar surface, whatever the reason, their absorption by the solar elements can result to a partial radiation pressure favoring thus heavier elements to escape from the Sun, e.g. oxygen compared to the hydrogen. Outstreaming solar axions, wherever their place of birth inside the sun, can be converted in a magnetic field **B** via the Primakoff effect (~$\mathbf{B}^2$-dependence), and, the created photons are outwards directed resulting to an also outwards pointing radiation pressure. In fact, the recent spectroscopic observations of a 50x30 $Mm^2$ solar surface [27] show clearly that the solar oxygen abundance in the pore (= tiny *sunspot* with up to a few kGauss magnetic field) and its neighborhood is considerably higher, reaching values up to a factor of ~3 larger than the average, and decreasing gradually moving away from it (see Figure 3). In general, magnetic concentrations exhibit higher abundances [27]. We notice here that the authors of ref. [27] conclude that "*this spatial variation of the solar oxygen abundance is likely an artifact of the imperfect modeling*" [27,28], since there is otherwise no physical reason to expect the actual abundance to exhibit spatial variations in the solar photosphere, indicating that important physical ingredients are still missing [27]. Interestingly, they conclude that this behavior is probably due to the presence of magnetic fields(!). In Figure 7 we plot the oxygen abundance as a function of the underlying magnetic field (kindly given to us by the authors of ref. [27]). Remarkably, as can be seen in Figure 7, the experimental points show also a $B^2$-dependence. Since there is not (yet) an alternative explanation for this, the celebrated axion scenario can be at the origin of this. In a future publication we will address this novel axion signature more in details [29]. However, we only recall already here that the invoked axion(-like?) scenario is probably not much different than the working principle of CAST phase I and/or II [1] as it was first suggested by Pierre Sikivie [30] and was extended by vanBibber-McIntyre-Morris-Raffelt for the case of a not perfectly empty magnetic space [18].

In favour of the axion-scenario, we notice that it has been observed that in most of the cases soft X-ray brightenings occur near the position of pores [31] (see also Figure 7a,f in ref. [31]). The not so close temporal correlation between the occurrence of transient brightenings and the observed formation of pores does not necessarily contradict the axion picture, since it can reflect the required fine tuning between axion restmass and local plasma frequency, the build-up of magnetic field strength/configuration, etc. [1]. In fact, unbiased from the axion scenario, Shimizu [32] concludes that "it is crucial to understand how magnetic fields involved in satellite spots [= pores] develop with occurrence of microflares … since observations indicate that emergence of magnetic flux probably plays a vital role in triggering soft X-rays transient brightenings", asking also: "what is the origin of such small energy releases in solar atmosphere?" , indicating the underlying mechanism is missing so far with conventional physics. We also would like to mention here that in earlier work [33] it was also pointed out that "pores were accompanied by highly localized soft X-rays". In addition, Parkinson [34] concluded in 1972 referring to pores and X-ray observations that 'considerable amounts of hot material, ≈4 MK, are being stored in the corona'.

Finally, combining all these kinds of signatures, one might be able to constrain the appropriate parameter values in direct axion searches with Earth-bound experiments. One first practical conclusion is that axion helioscopes should lower their threshold energy below a few 100 eV, following the scenarios with (a few 100 eV) massive and/or light axions as being co-responsible for the dominant and otherwise unexplained X-ray emission in this energy range, from the quiet and the active Sun alike. The axion signatures discussed remained unnoticed before, probably because of their multifaced appearance like QCD-inspired axions, massive Kaluza–Klein axions, or other exotic forms which have not yet been predicted.

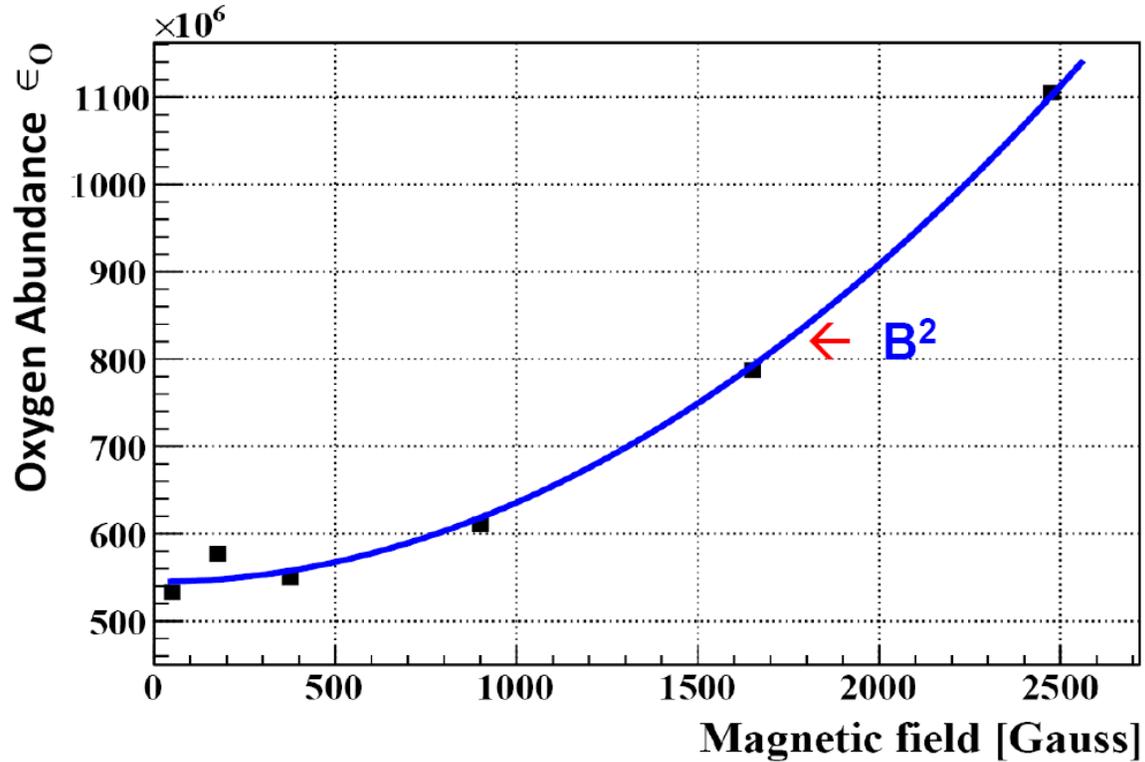

**Figure 7.** Solar oxygen abundance $\varepsilon_0$ as a function of the magnetic field **B** at the base of the photosphere near a pore. The blue line shows the $B^2$-dependence (including a constant component) through the experimental points (Courtesy Hector Socas-Navarro) [27]. The fit value for the $B^x$-term gives x=2.12±0.27.

**Acknowledgements**
We would like to thank Hector Socas-Navarro for the informative email exchange we had on questions regarding their oxygen abundance measurements, and in particular for giving us the experimental values, which allowed us to derive Figure 7 of this work. We also thank Susan Leech O'Neale for reading the manuscript. All kind of support we have received, from the greek funding agency GSRT and the Physics Department of the University of Patras, is gratefully acknowledged. This research was partially supported by the ILIAS (Integrated Large Infrastructures for Astroparticle Science) project funded by the EU under contract EU-RII3-CT-2004-506222.